\documentstyle[12pt]{article}
\begin{document}
\title{ Anomalies and Decoupling of charginos and neutralinos 
in the MSSM }
\author{ J. Lorenzo Diaz-Cruz  \\
  Instituto de Fisica, BUAP,  \\
 Ap. Postal J-48, 72500 Puebla, Pue., Mexico \\
      }
\maketitle
\begin{abstract}
We study the contribution of charginos and neutralinos 
of the Minimal SUSY extension of the Standard model (MSSM)
to the 1-loop vertices $ZAA$, $ZZA$, $ZZZ$, 
and examine the related cancellation of anomalies.
It is found that when the SUSY parameter $\mu$ 
satisfies $|\mu| >> M,M', m_W$, the couplings 
of charginos and neutralinos with the gauge bosons
become purely vectorial, 
and then their contribution to the amplitudes for 
$ZAA$, $ZZA$ and $ZZZ$ 
vanish, which implies that this sector of 
the MSSM does not generate a Wess-Zumino term.
We evaluate also the contribution of
charginos and neutralinos to the $\rho$ parameter,
and find that $\rho=0$ in the large-$\mu$ limit.

\end{abstract}

\newpage

\normalsize


The minimal SUSY standard model (MSSM) \cite{susyrev,sugrarev} 
has become one the most preferred extension of the 
standard model (SM). 
The success of the MSSM is not only due to its ability 
to imitate the SM agreement with most high energy data, 
but also because it gives a plausible explanation
of the new results that seem to be in conflict with
the SM (e.g. the large $Z\to b\bar{b}$ width \cite{zbbsusy},
the $ee\gamma\gamma$ event observed at FNAL \cite{kanetal}).
Moreover, the model predicts new signatures 
associated to the superpartners that are expected
to appear in the future colliders (LHC,NLC), or 
even in the present ones (FNAL/LEP).  
However, if the superpartners are not light,
it will become relevant to search for
any indirect physical effect that could be left
by them, and to verify by explicit 
calculations their expected decoupling.

Since gauge invariant
masses are allowed in vector-like theories, 
the effects of heavy fermion decouple 
from the corresponding low-energy 
effective lagrangian \cite{decoth}. 
However, in chiral theories heavy particles
do not decouple in general. For instance, 
if the Higgs mechanism of spontaneous
symmetry breaking (SSB) is used to generate masses, 
the associated v.e.v. ($v$) is fixed by the scale
of the interactions, then in order to generate a heavy 
fermion mass ($m >>v$), a large Yukawa coupling is required, 
which induces strong effects that prevent 
the decoupling of the fermion.
Chiral fermions may also pose another problem, namely 
the appearance of anomalies in gauge currents.
An interesting problem arises when the anomaly
cancellation occurs between fermions with very 
different masses \cite{veltman,dhoker}. In this case, 
integrating out the heavy fermion 
leaves an anomalous effective theory, 
which is the signal of its non-decoupling \cite{preskill}.

In the MSSM there are two sources of anomalies:
the ones due to quarks and leptons, and the
ones due to the fermionic partners of 
the Higgs bosons, the higgsinos, which 
cancel separately to make
the model anomaly-free. 
In fact, the higgsinos are not  mass eigenstates, 
they mix with the superpartners of the
gauge bosons (gauginos), 
and the resulting charged and neutral eigenstates are 
known as chargino and neutralino, respectively.
Gauginos do not contribute to the 
anomaly because their couplings to the gauge bosons 
are vectorial.

In this letter we are interested in studying the 
effects of heavy charginos and neutralinos, 
and to understand the role that the anomalies 
can play for their decoupling. In particular, we evaluate 
the contribution of charginos and neutralinos 
to the vertex $ZAA$, $ZZA$, $ZZZ$
and to the $\rho$ parameter, focusing in the
limit when the SUSY parameter $\mu$ satisfies
$|\mu| >> M,M',m_W$, where $M,M'$ denote the
gaugino soft-breaking masses, and the W-boson
mass $m_W$ is used to
characterize the electroweak scale.
\footnote{The case when a complete
supermultiplet is integrated out was
discussed in \cite{ssanom}, however in this paper 
we are interested in the limit when 
only the superpartners are heavy.}

Although the MSSM is anomaly-free, it is 
relevant to understand the conditions under 
which charginos and neutralinos participate 
in the cancellation of anomalies, since 
this can play an important role for their 
decoupling. For instance, if it were possible 
for a heavy higgsino to contribute to the anomaly,
the cancellation of anomalies would take place between 
different scales, which could prevent its decoupling.

In order to identify the origin of anomalies we shall  
work with 4-component spinors for the 
gaugino and higgsino fields, 
however, the analysis of the large-$\mu$ limit
and the calculations of interest will be performed in the 
mass-eigenstate basis, namely in terms of charginos 
and neutralinos. 
We shall review first the lagrangian of the model, 
focusing mainly in the interaction of 
the gauge bosons ($A_{\mu},W^{\pm}_{\mu},Z_{\mu}$), 
with the charged 
($\tilde{H}$) and  neutral ($\tilde{H}_1, \tilde{H}_2$) 
higgsinos, and with the wino ($\tilde{W}$),
photino ($\tilde{A}$) and zino ($\tilde{Z}$) fields.

The lagrangian for the mass and mixing
terms of gauginos and higgsinos in the MSSM is given by:
\begin{eqnarray}
{\cal{L}}&=&M_{\tilde{W}} \bar{\tilde{W}}\tilde{W}+ 
   \frac{M_{\tilde{A}}}{2} \bar{\tilde{A}}\tilde{A}+ 
   \frac{M_{\tilde{Z}}}{2} \bar{\tilde{Z}}\tilde{Z}+ 
   \mu \bar{\tilde{H}}\tilde{H} \nonumber \\
 & & + \frac{M_{\tilde{Z}}-M_{\tilde{A}}}{2} \tan2\theta_W
     \bar{\tilde{A}}\tilde{Z} 
      -\frac{\mu}{2}[ \bar{\tilde{H_1}}\tilde{H_2}+ 
       \bar{\tilde{H_2}}\tilde{H_1}] \nonumber \\
  & & -\frac{g}{\sqrt{2}}[\bar{\tilde{Z}}P_R \tilde{H_1} H^1_1-
        \bar{\tilde{H_2}}P_R \tilde{Z} H^2_2 + h.c.] \nonumber \\
  & & -g[\bar{\tilde{W}}P_R \tilde{H} H^1_1+
        \bar{\tilde{H}}P_R \tilde{W} H^2_2 + h.c.]
\end{eqnarray}
which includes the interaction of gauginos and higgsinos with
the neutral components of the scalar Higgs doublets 
($H^1_1,H^2_2$).
The zino and photino masses can be expressed
in terms of the  soft-breaking masses  $M$ and $M'$,
as follows:
$ M_{\tilde{Z}}=M\cos^2\theta_W+M'\sin^2\theta_W$,
$ M_{\tilde{A}}=M\sin^2\theta_W+M'\cos^2\theta_W$.

After SSB the Higgs scalars acquire v.e.v.'s 
($<H^1_1>=v_1$ and $<H^2_2>=v_2$), and
the trilinear terms in eq. (1) generate a mixing among 
the gauginos and higgsinos. The resulting 
mass-mixing matrices need to be 
diagonalized; the mass-eigenstates and the 
diagonalizing matrices depend in general
on the parameters $M, M', \mu$ and $\tan\beta \, (=v_2/v_1)$.
The diagonalizing matrices ($U,V$) for the charged case 
can be found in \cite{gunhaba}, whereas the $(4\times 4)$
matrix corresponding to the neutral case ($Z$)
is evaluated numerically. 
The charginos and neutralinos are denoted by:
$\tilde{\chi}^+_i \; (i=1,2)$ and 
$\tilde{\chi}^0_j \; (j=1-4)$, 
respectively.

The interaction of gauginos and higgsinos 
with the gauge bosons of the model
are described by the lagrangian: 
\begin{eqnarray}
{\cal{L}} &=& e [\bar{\tilde{W}}\gamma_{\mu} \tilde{W} 
   + \bar{\tilde{H}}\gamma_{\mu} \tilde{H}] A^{\mu}
   - e [\bar{\tilde{A}}\gamma^{\mu} \tilde{W} W^+_{\mu}
   + \bar{\tilde{W}}\gamma^{\mu} \tilde{A} W^-_{\mu}] \nonumber \\
  & & - g\cos\theta_W [ \bar{\tilde{Z}}\gamma^{\mu} \tilde{W} W^-_{\mu} 
    + \bar{\tilde{W}}\gamma^{\mu} \tilde{Z} W^+_{\mu} 
    - \bar{\tilde{W}}\gamma_{\mu} \tilde{W} Z^{\mu}] \nonumber \\ 
  & & + \frac{g}{2cos\theta_W}  
     [\cos2\theta_W \bar{\tilde{H}}\gamma_{\mu} \tilde{H} 
-\frac{1}{2}( \bar{\tilde{H_1}}\gamma_{\mu} \gamma_5\tilde{H_1}     
    -\bar{\tilde{H_2}}\gamma_{\mu} \gamma_5\tilde{H_2}) ]Z^{\mu}   
\end{eqnarray}

We can discuss now the origin of anomalies in the higgsino
sector. Before SSB the charged higssino couplings to the 
neutral gauge bosons are of vector type, and at this stage 
it does not contribute to the gauge anomaly. 
However, after SSB the mixing 
treats in a different way the L-  and
R-handed components of the charged higgsinos and winos,
which induces an axial-vector part for their couplings, 
then each chargino contributes to the anomaly,
but with opposite signs for the MSSM 
to remain anomaly-free. However,
the coupling becomes again vector-like 
for $tan\beta=1$ $(v_1=v_2)$. 
On the other hand, the neutral higgsinos 
contribute to the anomaly, because their 
couplings have an axial-vector part even before SSB.
However, these axial-vector couplings also vanish
in the large $\mu$ limit.

The complete Feynman rules for the interaction of the charginos 
and neutralinos are summarized in \cite{susyrev,gunhaba,hhunt}.
For the purpose of evaluating the 3-point vertex functions 
$Z_{\mu}A_{\nu}A_{\rho},Z_{\mu}Z_{\nu}A_{\rho}$
and $Z_{\mu}Z_{\nu}Z_{\rho}$, we only need to specify the 
following vertices: 

\begin{eqnarray}
\chi^+_i\chi^-_i A^{\mu}&: & -ie \gamma^{\mu}   \\
\chi^+_i\chi^0_j W^{-\mu}&: & +ig \gamma^{\mu}
                  (O^L_{ij}P_L+O^R_{ij}P_R)   \\
\chi^+_i\chi^-_j Z^{\mu}&: &  -i\frac{g}{cos\theta_W} 
        \gamma^{\mu}(O'^L_{ij}P_L+O'^R_{ij}P_R)    \\
\chi^0_i\chi^0_j Z^{\mu}&: &  -i\frac{g}{cos\theta_W} 
        \gamma^{\mu}(O''^L_{ij}P_L+O''^R_{ij}P_R) 
\end{eqnarray}
where $P_{R,L}=(1 \pm \gamma_5)/2$, and:
\begin{eqnarray}
O^L_{ij}&=& Z_{i2}V^*_{j1}-\frac{1}{\sqrt{2}}Z_{i4}V^*_{j2} \\
O^R_{ij}&=& Z^*_{i2}U_{j1}+\frac{1}{\sqrt{2}}Z^*_{i3}U_{j2} \\
O'^L_{ij}&=& -V_{i1}V^*_{j1}-\frac{1}{2}V_{i2}V^*_{j2} +
          \delta_{ij} sin^2 \theta_W  \\
O'^R_{ij}&=& -U_{j1}V^*_{i1}-\frac{1}{2}U_{j2}U^*_{i2} +
          \delta_{ij} sin^2 \theta_W   \\
O''^L_{ij}&=& \frac{1}{2}(Z_{i4}Z^*_{j4}-Z_{i3}Z^*_{j3}) \\
O''^R_{ij}&=& -\frac{1}{2}(Z_{i4}Z^*_{j4}-Z_{i3}Z^*_{j3}) 
\end{eqnarray}

 We can evaluate now the SUSY contribution to the 
1-loop vertex $Z_{\mu}A_{\nu}A_{\rho}$.
 Using the CP-properties of the vector (V) and axial-vector (A)
currents, it can be shown that the 3-point functions
$VVV$ and $VAA$ vanish in general. These CP-properties can be
used also to show that sfermions, goldstone, Higgs and gauge bosons 
do not contribute to the vertex $Z_{\mu}A_{\nu}A_{\rho}$. 
Thus, the amplitude can only arise from the triangle 
graphs with charged fermions inside the loop.
The contribution from each chargino ($\tilde{\chi}^+_i$)
to the amplitude can for the vertex 
$Z_{\mu}(q)A_{\nu}(k_1)A_{\rho}(k_2)$ 
can be obtained from the results of \cite{barroso}, 
and it is written as follows,
\begin{eqnarray}
T^{ \mu \nu \rho}_i&=& \frac{4\alpha g a'_{ii}Q^2_i }{\pi cos\theta_W}
\{ \varepsilon^{\mu \nu \rho \alpha}
     [k_{1\alpha}  (f_1 k^2_2-f_3 k_1.k_2 )-
      k_{2\alpha}  (f_2 k^2_1-f_3 k_1.k_2 ) ]  +  \nonumber \\
& & \varepsilon^{\alpha \mu \beta \nu }k_{1\alpha}k_{2\beta}
      [f_2 k^{\rho}_1+f_3 k^{\rho}_2]+
\varepsilon^{\alpha \mu \beta \rho }k_{1\alpha}k_{2\beta}
      [f_1 k^{\nu}_2+f_3 k^{\nu}_1] \},
\end{eqnarray}
where the axial coupling is given by: 
$a'_{ij}=O'^R_{ij}-O'^L_{ij}$, and the functions $f_i$ are  
defined by:
\begin{eqnarray}
f_1&=& \int^{1}_0 dx_1  \int^{1-x_1}_0 dx_2 \frac{x_1(x_1-1)}{D}, \\ 
f_2&=& \int^{1}_0 dx_1  \int^{1-x_1}_0 dx_2 \frac{x_2(x_2-1)}{D}, \\ 
f_3&=& \int^{1}_0 dx_1  \int^{1-x_1}_0 dx_2 \frac{x_1x_2}{D}, 
\end{eqnarray}                                        
with $D=M^2_{\chi^+_i}+ x_2(x_2-1)k^2_1 + x_1(x_1-1) k^2_2
-2k_1.k_2 x_1x_2$.
Then, the condition for the non-conservation of
the axial-current is written as:
\begin{eqnarray}
q_{\mu}T^{\mu \nu \rho}&=&\sum_i \frac{4\alpha g a_{ii}Q^2_i }
       {\pi cos\theta_W}
\varepsilon^{\mu \nu \rho \alpha} k_{1\alpha} k_{2\mu}  
  (f_1 k^2_2+f_2 k^2_1-2f_3 k_1.k_2 ) \nonumber\\  
&=& \sum_i \frac{4\alpha g a_{ii}Q^2_i }{\pi cos\theta_W}
\varepsilon^{\mu \nu \rho \alpha} k_{1\alpha} k_{2\mu}   
[\frac{1}{2}-M^2_{\chi^+_i}f_0 ]
\end{eqnarray}
where
\begin{equation}
f_0=\int^{1}_0 dx_1  \int^{1-x_1}_0 dx_2 \frac{1}{D} , 
\end{equation}

When both charginos are taken into account the
mass independent term (i.e. the anomaly)
should cancel ($\Sigma_i q_{\mu}T^{ \mu \nu \rho}_i=0$), 
as we have verified by the direct substitution of the elements
of $U,V$ in eq. (17), namely: $\sum_i Q^2_ia'_{ii}=0$.

In order to study the limit when the mass parameters 
are very large ($>>m_W$), one can use the 
results of \cite{gunhabb}, which presents an analytical
expression for the diagonalizing matrices $U,V,Z$, 
under the assumption that the couplings are CP-invariant,
and with the mass parameters satisfying
the conditions: $ |M\pm \mu|, |M' \pm \mu |>> m_W$, 
and $|M\mu |>> m^2_W\sin2\beta$. Then, the 
matrices $U,V$ are given by: 

\begin{equation}
U = \left(
\begin{array}{ll}
\; 1  &  \sqrt{2} m_W\frac{ Mc_{\beta}+\mu s_{\beta}}{M^2-\mu^2}  \\
 - \sqrt{2} m_W\frac{ Mc_{\beta}+\mu s_{\beta}}{M^2-\mu^2}   & \; 1  
\end{array}
\right)
\end{equation}

\begin{equation}
V= \left(
\begin{array}{ll}
 \; 1    &  \sqrt{2}m_W\frac{ Ms_{\beta}+\mu c_{\beta}}{M^2-\mu^2} \\
 -\sqrt{2} m_W\frac{ Ms_{\beta}+\mu c_{\beta}}{M^2-\mu^2} & \; sign(\mu)  
\end{array}
\right),
\end{equation}
whereas the neutralino Z matrix  
takes the form:

\begin{equation}
Z= \left(
\begin{array}{ll}
 A & B \\
 C & D
\end{array}
\right)
\end{equation}
with: 
 
\begin{equation}
A= \left(
\begin{array}{ll}
 \; \; 1 & - \frac{ m^2_Zs_{2W}(M'+\mu s_{2\beta})}{2(M'-M)(M'^2-\mu^2)}\\
 -\frac{ m^2_Zs_{2W}(M+\mu s_{2\beta})}{2(M'-M)(M'^2-\mu^2)} & \;\; 1
\end{array}
\right)
\end{equation}

\begin{equation}
B= \left(
\begin{array}{ll}
  \frac{- m_Zs_{W}(M'c_{\beta}+\mu s_{\beta})}{M'^2-\mu^2}    
 &  \frac{ m_Zs_{W}(M'c_{\beta}+\mu s_{\beta})}{M'^2-\mu^2}   \\
  \frac{ m_Zc_{W}(Mc_{\beta}+\mu s_{\beta})}{M^2-\mu^2}    
  & \frac{- m_Zc_{W}(Ms_{\beta}+\mu c_{\beta})}{M^2-\mu^2}    
\end{array}           
\right)
\end{equation}
 
\begin{equation}
C= \left(
\begin{array}{ll}
  \frac{- m_Zs_{W}(s_{\beta}- c_{\beta})}{\sqrt{2}(M'+\mu)}    
 &  \frac{ m_Zc_{W}(s_{\beta}- c_{\beta})}{\sqrt{2}(M+\mu) }     \\
  \frac{ m_Zs_{W}(s_{\beta}+ c_{\beta})}{\sqrt{2}(M'-\mu)}  
  & - \frac{ m_Zc_{W}(s_{\beta}+ c_{\beta})}{\sqrt{2}(M-\mu) } 
\end{array}
\right)
\end{equation}

\begin{equation}
D= \left(
\begin{array}{ll}
  \frac{1}{\sqrt{2}}  & \frac{1}{\sqrt{2}} \\
  \frac{1}{\sqrt{2}}  & -\frac{1}{\sqrt{2}}
\end{array}
\right)
\end{equation}
and where: $s_W= \sin \theta_W, s_{2W}= \sin 2\theta_W$,  
$s_\beta=\sin \beta, c_\beta=\cos \beta, 
s_{2\beta}=\sin 2 \beta $.

Our results will be presented assuming also 
that $|\mu| >> M,M',m_W$,  and keeping 
only the leading terms in $1/\mu$, in whose case the 
mass eigenstates  are given by:
$M_{\chi^+_{2}}= \mu$, $M_{\chi^+_{1}}= M$, 
$M_{\chi^0_{1}}= M'$, $M_{\chi^0_{2}}= M$, 
$M_{\chi^0_{3}}= \mu$, $M_{\chi^0_{4}}= \mu$, 
The chargino mixing matrices take the values
$U=V=1$, whereas the expression for $Z$ 
also reduces considerably. 
In this limit one obtains
$(M^2_{\chi^+_i}/D) \to 1$, and consequently 
$(M^2_{\chi^+_i}f_0) \to 1/2$, which appears as if an
anomalous term would remain.
However, a careful analysis of the couplings 
shows that in this limit the lightest chargino 
becomes a pure gaugino ($\tilde{\chi}^+_1=\tilde{W}$),
which does not have axial couplings, i.e. $a'_{11}=0$, 
and the heavy chargino becomes a pure higgsino,
which also has $a'_{22}= 0$, thus 
$T^{ \mu \nu \rho} =0$, and charginos decouple
from this function.

Similarly, we can evaluate the contribution of
charginos to the vertex $Z(q_1)Z(q_2)A(k)$,
which we denote by $R^{\mu \nu \rho}$.
In this case the condition for
the conservation of the axial-vector current 
is written as:
\begin{eqnarray}
q_{\mu}R^{\mu \nu \rho}&=&-\sum_i \frac{4\alpha g a'_{ii}v'_{ii}Q^2_i }
{\pi cos^2\theta_W} \varepsilon^{\mu \nu \rho \alpha} q_{2\alpha} k_{\mu}  
\end{eqnarray}
which must be zero because of the anomaly cancellation.
The vector coupling is given by $v'_{ij}=O'^R_{ij}+O'^L_{ij}$.
Moreover, the vertex $R^{\mu \nu \rho}$ itself
also vanishes in the large $\mu$ limit,
because the chargino couplings $a'_{ij}$ vanishes. 

The amplitude for the 1-loop vertex $Z(q_1)Z(q_2)Z(q_3)$ 
($=S^{\mu \nu \rho}$), receives contributions from charginos 
and neutralinos, and the conservation of the currents
takes the form:
\begin{eqnarray}
q_{\mu}S^{\mu \nu \rho}&=&-\sum_{ij} \frac{g^3}{2\pi^2 cos^3\theta_W} 
(v''^2_{ii}+\frac{1}{3}a''^2_{ii})a''_{ii} 
\varepsilon^{\mu \nu \rho \alpha} q_{2\alpha} q_{3\mu}  
\end{eqnarray}
which also vanishes because of the anomaly cancellation.
Charginos do not contribute to the 
the vertex $S^{\mu \nu \rho}$ itself,
because their axial-vector coupling vanish
in the limit $\mu >>M,M',m_W$, and 
the 3-point function with vector couplings only
($VVV$) vanishes.
Moreover, since the axial-vector couplings of the neutralinos
vanish in the large-$\mu$ limit, 
its contribution to $S^{\mu \nu \rho}$ itself
vanishes too.
Thus, integrating out the charginos and neutralinos
does not leave a mass-independent Wess-Zumino
term in the low-energy effective lagrangian.

Another process that illustrates this decoupling result 
is the (1-loop) Higgs amplitude $hA_{\mu}A_{\nu}$. 
For the light Higgs scalar ($h^0$) of the MSSM, 
the contribution of charginos to the
amplitude is proportional to the function \cite{hhunt}:
\begin{equation}
I(m_h,m_{\chi^+_i})= \frac{4m_W m_{\chi^+_i}A_{ii}}{sin\beta}
\int
^1_0  dx \int^{1-x}_0 dy \frac{1-4xy}{ m^2_{\chi^+_i}-xym^2_h}
\end{equation} 
where $A_{ij}$ is a dimensionless coefficient that depends
on the elements of the matrices $U,V$.
The amplitude behaves like $m_W/M_{\chi^+_i}$, 
and it vanishes in the large $\mu$ limit. 
On the other hand, the contribution of the top quark
to the amplitude becomes a constant in the
limit of a large top mass
\footnote{In order to derive systematically the full
effective lagrangian that remains 
after the charginos/neutralinos are integrated out, 
one could use the method of ref. \cite{aitch}, 
as it was done in \cite{yaoetal} for the 
integration of the top quark in the SM, however the 
specific cases discussed in the present paper,
allows us to understand the limit of
heavy charginos and neutralinos.}.

Finally, we evaluate the contributions
of heavy charginos and neutralinos to 
Veltman's $\rho$ parameter.
Complete calculations of radiative corrections for the
MSSM have been performed in the literature \cite{hollik}, 
however the results for the chargino/neutralino sector
are presented only in numerical form, which does not help 
to clarify the subtleties associated
with the heavy-mass limits.

The definitions of the $\rho$ parameter in terms 
of the self-energies of the gauge bosons
($\Pi_{ZZ},\Pi_{WW}$) is the following \cite{pestak}:
\begin{eqnarray}
\rho &=& \frac{\Pi_{ZZ}(0)}{m^2_Z}- 
               \frac{\Pi_{WW}(0)}{m^2_W} 
\end{eqnarray}
where $\Pi(0)'s$ are obtained from the expansions 
$\Pi_{ij}(q^2)=\Pi_{ij}(0)+\Pi'_{ij}(0)q^2$.

The total contribution of chargino and neutralino 
to the self-energies, keeping only the leading 
terms in the limit $|\mu| >> M,M', m_W$, is: 
\begin{eqnarray}
{\Pi}_{WW}(0)&=&\frac{g^2}{8\pi^2} [\mu^2+A_0(\mu)] \\
{\Pi}_{ZZ}(0)&=& \frac{g^2}{8\pi^2 c^2_W} [\mu^2+A_0(\mu)] 
\end{eqnarray}
where $A_0(\mu)=-\mu^2+2\mu^2\log(\mu/\mu_0)$, and
$\mu_0$ is the mass-scale that  arises in the 
$\bar{MS}$-scheme with dimensional regularization.

Thus, it follows eqs.  (39) and (31) that 
$\rho=0$. 
This result can be understood if we remind
that $\rho$ is associated with the breaking of 
isospin, and
the large-$\mu$ limit does not induce a
mass-splitting among the charginos and neutralinos 
that interact with the gauge bosons, 
i.e. $M_{\chi^0_{3,4}}=M_{\chi^+_2}$; 
thus $\rho$ must vanish in this limit.

In conclusion, we have studied the 
effects that remain after taking
the large $\mu$ limit in the MSSM.
It is found that despite the fact that charginos and
neutralinos contribute to the anomaly, they
do not induce a Wess-Zumino terms in the effective lagrangian,
unlike other cases studied in the literature 
\cite{rball}. This result can be explained by
the form of the soft-SUSY breaking mass terms, 
which do not allow a large mass-splitting among 
the higgsinos; moreover, the large-$\mu$ limit 
in the gaugino-higsino sector of the MSSM is obtained 
by rendering large a dimensional parameter,
associated to gauge invariant mass terms,
which does not produce strong interaction
effects. A large mass splitting among higgsinos
could be obtained if it were possible to include a 
mass term for each higgsino in the lagrangian, however 
this type of term is not soft \cite{susyrev}.  Moreover,
we also found that the effects of charginos-neutralinos 
to the $\rho$ parameter vanish
in the large $\mu$ limit. Thus, they decouple
in all quantities studied in this paper.

We have also reviewed the pattern of anomaly 
cancellation due to higgsinos in the Minimal SUSY extension 
of the Standard model (MSSM), and found that they 
have different characteristics as compared with the 
ones due to quarks and leptons. For instance, 
when $tan\beta=1$ it happens that
the charged higgsinos do not contribute to the
anomaly, whereas the neutralinos do not contribute 
to the anomaly when $\mu$ is large. 
This result suggest an alternative 
mechanism to obtain anomaly-free theories. 
In the usual approach, it is assumed that the chirality 
of the fermions is fixed, then the representations
are adjusted in order to cancel the anomalies.
However, in extended SUSY models, new particles 
are predicted, whose chiralities are not 
known yet, and if their fixing depends on
some unknown parameters, then it may be possible that 
those new parameters have values that make the 
theory anomaly-free.

{{\bf Acknowledgment.-} Valuable discussions with G. Kane, 
M.J. Herrero, J.J. Toscano, M. Hernandez and correspondence 
with E. D'Hoker are acknowledged. 
This work was supported by CONACYT and SNI (M\'exico).}
\newpage



\begin{thebibliography}{9}
\bibitem{susyrev}  For a review see: 
H. Haber and G.L. Kane, Phys. Rep. 117 (1985) 75.
H. Haber, "Introductory low-energy superstmmetry", 
lectures given at TASI-92,
U. of Colorado, SCIPP-92/93.
\bibitem{sugrarev} H.P. Nilles, Phys. Rep. 110 (1984) 1.
\bibitem{zbbsusy} J. Wells and G.L. Kane, 
Phys. Rev. Lett. 76 (1996) 869, and references therein.
\bibitem{kanetal} S. Ambrosiano et al., 
Phys. Rev. Lett. 76 (1996) 3498.
\bibitem{decoth} T. Appelquist and Carazzone, 
Phys. Rev. D11 (1975) 2856.
\bibitem{veltman} T. Stirling and M. Veltman,
Nucl. Phys. B189 (1981) 577.
\bibitem{dhoker}  E. Farhi and E. D'Hoker,
Nucl. Phys. B248 (1984) 59,77.
\bibitem{preskill} J. Preskill, Ann. Phys. (N.Y.) 210 (1991) 323.
\bibitem{ssanom} S. Ferrara et al., Nucl. Phys. B417 (1994) 238.
\bibitem{gunhaba} J.F. Gunion and H. Haber, 
Nucl. Phys. B272 (1986) 1; J. Rosiek,
Phys. Rev. D41 (1990) 3464.    
\bibitem{hhunt} S. Dawson et al., {\it The Higgs Hunter Guide},
(Adison Wesley, 1992).
\bibitem{barroso} A. Barroso et al., Z. Phys. C33 (1986) 243.
\bibitem{gunhabb} J.F. Gunion and H. Haber, 
Phys. Rev. D37 (1988) 2515.
\bibitem{aitch} I.J.R. Aitchison and C.M. Fraser,
Phys. Rev. D31 (1985) 2605.
\bibitem{yaoetal} Guy-lin Liu, H. Steger and Y.P. Yao,
Phys. Rev. D44 (1991) 2139.    
\bibitem{hollik} W. de Boer et al., preprint IEKP-KA/96-08;
see also the second ref. in [1].
\bibitem{pestak} M. Peskin and T. Takeuchi,
Phys. Rev. D46 (1992) 381.
\bibitem{rball} R. Ball, Phys. Rep. 182 (1989) 1.


\end{thebibliography}
\end{document}